\documentclass[conference]{IEEEtran}
\usepackage{tikz}
\usetikzlibrary{positioning, arrows.meta, fit, calc, shapes.geometric, backgrounds}

\usepackage{caption}
\usepackage{newtxtext}
\usepackage{newtxmath}
\usepackage{cite}
\usepackage{amsmath,amsfonts}%amssymb
\usepackage{graphicx}
\usepackage{textcomp}
\usepackage{xcolor}
\usepackage{booktabs}
\usepackage{multirow}
\usepackage{subcaption}
\usepackage{array}
\usepackage{url}
\usepackage{bm}
\usepackage{balance}
\usepackage{algorithm}
\usepackage{algorithmic}
\usepackage{float}

\usepackage[bookmarks=false]{hyperref}

\begin{document}
\captionsetup[figure]{labelfont={rm},labelformat={default},labelsep=period,name={Fig.}}

\title{Optimizing Tracking Accuracy in Energy-Constrained Multimodal ISAC via Lyapunov-Driven Heterogeneous Mixture-of-Experts}

\author{
    \IEEEauthorblockN{
        Wenqi Fan\IEEEauthorrefmark{1}, 
        Ning Wei\IEEEauthorrefmark{1}, 
        Ahmad Bazzi\IEEEauthorrefmark{2},
        Rongyan Xi\IEEEauthorrefmark{3}, 
        Zhixian Song\IEEEauthorrefmark{1}\\
        You Li\IEEEauthorrefmark{4},
        Zhihan Zeng\IEEEauthorrefmark{4},  
        Yue Xiu\IEEEauthorrefmark{1},
        Chadi Assi\IEEEauthorrefmark{5}  
    }
\IEEEauthorblockA{
        \IEEEauthorrefmark{1}University of Electronic Science and Technology of China, Chengdu, China \\
        % Email: fwq897032532@gmail.com, wn@uestc.edu.cn, xiuyue12345678@163.com
    }
    \IEEEauthorblockA{
        \IEEEauthorrefmark{2}Wireless Research Lab, Engineering Division, New York University Abu Dhabi
(NYUAD), UAE\\
        % Email: ahmad.bazzi@nyu.edu
    }
\IEEEauthorblockA{
        \IEEEauthorrefmark{3}China Mobile Research Institute, Beijing, China \\
        % Email: \{xirongyan, dongjing, jinjing\}@chinamobile.com
    }
    \IEEEauthorblockA{
        \IEEEauthorrefmark{4}Southwest China Research Institute of Electronic Equipment (SWIEE), Chengdu, China\\
    }
    \IEEEauthorblockA{
        \IEEEauthorrefmark{5}Institute for Information Systems Engineering, Concordia University, Quebec, Canada \\
        % Email: Assi@ciise.concordia.ca
    }
}

\maketitle

\begin{abstract}
The integration of multimodal sensing and millimeter-wave (mmWave) communications is a key enabler for highly mobile vehicle-to-infrastructure (V2I) networks. However, continuous high-resolution visual sensing incurs prohibitive computational energy, while delayed sensing information causes severe beam misalignment. This paper establishes a physics-aware multimodal integrated sensing and communication (M-ISAC) framework that mathematically bridges network-layer queuing delays with physical-layer spatial uncertainty via the semantic age of information (AoI). Guided by this relationship, we aim to strike an optimal trade-off between the tracking posterior Cram\'er-Rao bound (PCRB) and system energy budgets, we formulate a stochastic mixed-integer non-linear programming (MINLP) problem. Addressing the coupled challenges of temporal computing congestion and non-convex constant modulus constraints, we propose a reinforcement learning (RL) framework empowered by a Lyapunov-driven heterogeneous mixture-of-experts (LD-H-MoE) architecture. By strictly decoupling temporal scheduling and spatial phase mapping into specialized subnetworks, the LD-H-MoE circumvents gradient conflicts prevalent in monolithic multi-task learning. Simulations demonstrate that the proposed LD-H-MoE achieves a highly-effective event-triggered sensing policy, yielding superior tracking accuracy and radio-frequency (RF) resilience while guaranteeing edge computing queue stability and long-term energy budgets.
\end{abstract}


\section{Introduction}
In 6G vehicle-to-infrastructure (V2I) networks, millimeter-wave (mmWave) and terahertz (THz) bands provide ultra-high data rates but require massive arrays and highly directional pencil beams to overcome severe path loss \cite{8999605, 7905941}. In high-mobility environments, these narrow beams are susceptible to spatial misalignment, turning frequent beam tracking into a high-overhead bottleneck that degrades reliable connectivity \cite{9246715}.

Multimodal integrated sensing and communication (M-ISAC) mitigates pure-radio vulnerabilities by fusing heterogeneous data \cite{10330577}. For example, radar tracks real-time motion, while cameras provide semantic context for angular alignment \cite{8114345}. However, multimodal fusion presents two critical challenges in edge networks. First, processing visual data incurs high energy consumption and queuing delays; thus, semantic age of information (AoI) must be incorporated to map data staleness to spatial uncertainty \cite{9380899,11418623,11353414,11373884,11355857,11346858,11316633,YU2024108842,10.1093/bib/bbae473}. Second, static fusion mechanisms fail to handle the non-convex constant modulus constraints of mmWave analog phase shifters \cite{10870338}.
Consequently, optimizing this M-ISAC system yields an intractable mixed-integer non-linear programming (MINLP) problem. Traditional optimization cannot jointly manage long-term AoI queue evolution and instantaneous phase mapping. While deep reinforcement learning (DRL) enables online decision-making without non-causal knowledge \cite{8714026}, standard monolithic DRL architectures suffer from multi-task gradient conflicts. 

To tackle these issues, we propose a novel Lyapunov-driven heterogeneous mixture-of-experts (LD-H-MoE) architecture \cite{fan2026heterogeneousmixtureofexpertsenergyefficientmultimodal}. Leveraging Lyapunov stochastic optimization, we decouple long-term energy and queue stability constraints into deterministic single-slot surrogate objectives \cite{10924670}. Guided by the drift-plus-penalty function, LD-H-MoE disentangles temporal resource allocation and spatial beamforming into specialized expert networks. The main contributions of this paper are summarized as follows:
\begin{itemize}
    \item We develop a cross-layer framework coupling multimodal semantic AoI with the posterior Cram\'er-Rao bound (PCRB). Recognizing that mmWave systems are predominantly vulnerable to spatial misalignment, we explicitly formulate the angular PCRB as our primary reliability metric, analytically mapping network-layer delays to physical-layer errors via a multivariate Taylor expansion.
    \item We integrate a virtual energy deficit queue---tracking the accumulated excess energy beyond predefined budgets---into the Lyapunov framework to strictly enforce long-term energy constraints.
    \item To tackle the hybrid discrete-continuous action space, we design an Actor-Critic architecture featuring decoupled temporal and spatial experts. Internally, a strict gradient isolation mechanism circumvents negative transfer, satisfying non-convex constant modulus constraints without degrading long-term scheduling memory.
    \item Simulations demonstrate that the proposed LD-H-MoE learns an efficient event-triggered sensing policy. It achieves dynamic modality-reliance, maintaining robust tracking accuracy in harsh environments while strictly adhering to long-term energy budgets.
\end{itemize}

\section{System Model and Problem Formulation}
We consider a highly mobile V2I scenario. The system consists of a single BS equipped with an $M$-element antenna array and a set of multimodal sensors (RGB camera and mmWave radar), serving $K$ high-mobility vehicles. The system operates in discrete time slots (symbols) $n \in \{1, \dots, N\}$.

\subsection{Communication and Channel Model}
The BS utilizes mmWave bands for downlink communication. Let $\mathbf{v}_k(n) \in \mathbb{C}^{M \times 1}$ denote the beamforming vector for the $k$-th vehicle at time slot $n$. The received signal at user $k$ is given by
\begin{equation}
y_k(n) = \mathbf{h}_k^H(n) \mathbf{v}_k(n) s_k(n) + \sum_{j \neq k} \mathbf{h}_k^H(n) \mathbf{v}_j(n) s_j(n) + z_k(n),
\end{equation}
where $s_k(n)$ is the transmitted data symbol intended for the $k$-th vehicle and $z_k(n) \sim \mathcal{CN}(0, \sigma^2)$ is the additive white Gaussian noise at the $k$-th vehicle. The geometric multipath channel $\mathbf{h}_k(n)$ is highly sensitive to the spatial state of the vehicle. In line-of-sight (LoS) dominated scenarios, it can be modeled as a multivariate function of the radial distance $d_k$ and the angle of arrival (AoA) $\theta_k$:
\begin{equation} \label{eq:channel_multivariate}
\mathbf{h}_k(d_k, \theta_k) = \alpha(d_k) e^{-j2\pi f_c \tau_k(d_k)} \mathbf{a}(\theta_k),
\end{equation}
where $f_c$ is the mmWave carrier frequency, $\alpha(d_k)$ is the large-scale path loss, $\tau_k(d_k)$ is the propagation delay, and $\mathbf{a}(\theta_k)$ is the array steering vector. The Doppler shift is omitted as it is compensated during preamble synchronization, and our focus remains on macroscopic beam misalignment over microscopic phase shifts.

\begin{figure}[htbp]
    \centering
    \includegraphics[width=0.9\columnwidth]{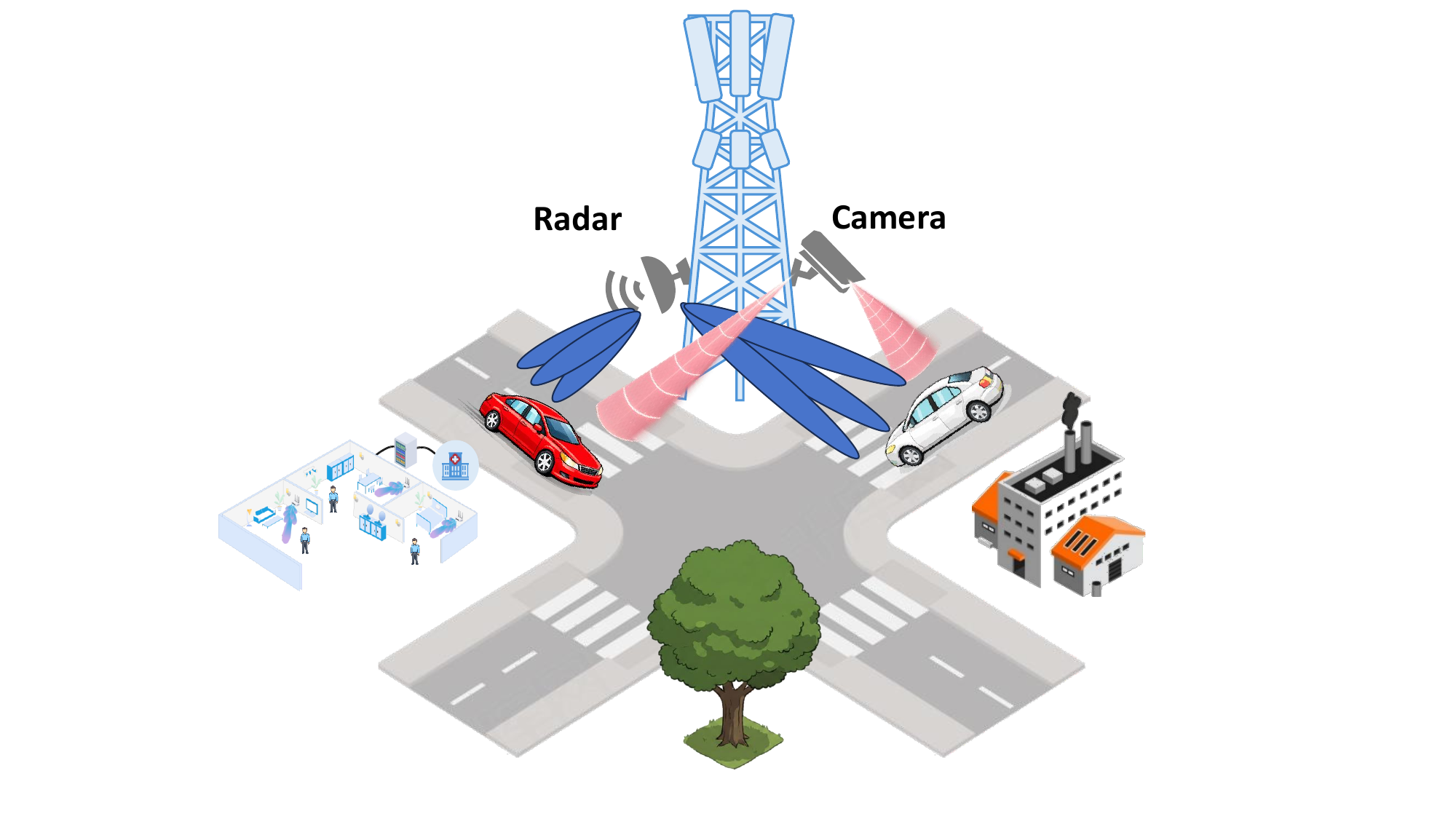}
    \caption{Illustration of the multimodal ISAC system in a V2I scenario.}
    \label{fig:system_scenario}
\end{figure}

\subsection{Edge Computing Queues and Dynamic Energy Model}
% Specifically, the heterogeneous sensing input comprises the visual data $S_{vis}(n)$ and the radar observation $\tilde{S}_{rad,k}(n)$. 
Specifically, the heterogeneous sensing input comprises the visual data and the radar observation. Let $\pi_k(n) \in \{0, 1\}$ denote the discrete sensor scheduling decision, where $\pi_k(n) = 1$ indicates that the visual camera is activated. Visual data processing is highly energy-intensive and constrained by the BS's edge computing capability. When $\pi_k(n) = 1$, a computational task of size $C_k$ (in CPU cycles) is generated.

To capture the time-coupled nature of the available computing resources, we establish a dynamic task queue $Q_k(n)$ for each vehicle $k$ at the BS. The evolution of the computing queue is formulated as \cite{10924670}:
\begin{equation} \label{eq:queue_evolution}
Q_k(n+1) = \max\left(0, Q_k(n) - f_k(n)\tau\right) + \pi_k(n)C_k,
\end{equation}
where $f_k(n)$ denotes the CPU computational frequency allocated to vehicle $k$, and $\tau$ is the slot duration. The total allocated frequency is strictly bounded by $\sum f_k(n) \le F_{max}$. Based on the dynamic voltage and frequency scaling (DVFS) model, the total system energy consumption $E_{total}(n)$ comprises the dynamic computational energy and the RF beam recovery penalty \cite{10870338,fan2026heterogeneousmixtureofexpertsenergyefficientmultimodal}:
\begin{equation} \label{eq:total_energy}
E_{total}(n) = \sum_{k=1}^K \kappa f_k(n)^3 \tau + \sum_{k=1}^K P_{misa, k}(n) \cdot E_{recovery},
\end{equation}
where $\kappa$ is the effective capacitance coefficient and $E_{recovery}$ is the RF energy overhead incurred by exhaustive beam sweeping due to beam misalignment probability $P_{misa,k}(n)$.

\subsection{Geometrically Coupled Vector AoI and Spatial Uncertainty}
Instead of utilizing a generalized scalar AoI, we propose a physics-aware vectorized AoI. Let $\mathbf{A}_k(n) = [A_k^{rad}(n), A_k^{tan}(n)]^T$ denote the semantic AoI. Since mmWave radar operates continuously with negligible delay, $A_k^{rad}(n+1) = 1$. The tangential AoI relies on the camera, and its evolution is strictly governed by the edge computing queue. Defining an indicator function $\mathbb{I}_k(n) \in \{0,1\}$ to signify whether the visual task in $Q_k(n)$ is completed, $A_k^{tan}(n)$ evolves as \cite{fan2026heterogeneousmixtureofexpertsenergyefficientmultimodal}:
\begin{equation} \label{eq:aoi_evolution}
A_k^{tan}(n+1) = 
\begin{cases} 
T_{queue,k}, & \text{if } \mathbb{I}_k(n) = 1 \\ 
A_k^{tan}(n) + 1, & \text{if } \mathbb{I}_k(n) = 0
\end{cases}
\end{equation}
The spatial uncertainty at the BS is modeled as a 2D Gaussian distribution, whose covariance matrix $\bm{\Sigma}_k(n) = \text{diag}\left( \sigma_{rad,k}^2(n), \sigma_{tan,k}^2(n) \right)$ is non-linearly coupled with the vector AoI \cite{7905941,9144087}:
\begin{equation}
\sigma_{rad,k}^2(n) \propto (u_k^{rad}(n) \cdot A_k^{rad}(n))^2 + \epsilon_{rad}^2,
\end{equation}
\begin{equation}
\sigma_{tan,k}^2(n) \propto (u_k^{tan}(n) \cdot A_k^{tan}(n))^2 + \epsilon_{cam}^2.
\end{equation}
where $u_k^{rad}(n)$ and $u_k^{tan}(n)$ denote the radial and tangential velocities of vehicle $k$ at time slot $n$, respectively. Due to the intermittent activation of the multimodal sensors, the BS relies on the delayed state estimate $[\hat{\theta}_k(n), \hat{d}_k(n)]^T$. Utilizing the first-order multivariate Taylor expansion of \eqref{eq:channel_multivariate}, and given that the radial and tangential movements are physically orthogonal, the AoI-driven channel error variance is formulated as:
\begin{equation} \label{eq:channel_error_variance}
\begin{split}
\mathbb{E}\left[\|\Delta \mathbf{h}_k(n)\|^2\right] &\approx \left\| \frac{\partial \mathbf{h}_k}{\partial \theta_k} \right\|^2 \frac{\sigma_{tan,k}^2(A_k^{tan}(n))}{d_k^2(n)} \\
&\quad + \left\| \frac{\partial \mathbf{h}_k}{\partial d_k} \right\|^2 \sigma_{rad,k}^2(A_k^{rad}(n)).
\end{split}
\end{equation}
Equation \eqref{eq:channel_error_variance} explicitly demonstrates the asymmetry of multimodal delays: the channel deterioration is intensely amplified by the tangential AoI at shorter distances within the Fraunhofer far-field boundary, enforcing the necessity of visual sensing to suppress angular uncertainty.

\subsection{Beam Misalignment Probability via the Q-Function}
It is worth noting that the angular uncertainty directly dictates the necessity of beam failure recovery, which heavily impacts the system energy $E_{total}(n)$. Specifically, a beam misalignment event occurs when the instantaneous angular error $|\Delta \theta_k(n)|$ exceeds half of the half-power beamwidth $\theta_{BW}/2$. Given that the angular error follows a zero-mean Gaussian distribution with variance $\sigma_{\theta,k}^2(n) = \sigma_{tan,k}^2(A_k^{tan}(n)) / d_k^2(n)$, the beam misalignment probability $P_{misa,k}(n)$ in \eqref{eq:total_energy} can be rigorously derived using the standard Q-function \cite{9246715}:
\begin{equation} \label{eq:pmisa_qfunction}
P_{misa,k}(n) = 2 Q\left( \frac{\theta_{BW} \cdot d_k(n)}{2 \sigma_{tan,k}(A_k^{tan}(n))} \right).
\end{equation}
Equation \eqref{eq:pmisa_qfunction} effectively couples the RF energy penalty with the edge computing dynamics.

\subsection{AoI-Coupled Posterior Cram\'er-Rao Bound (PCRB)}
To strictly guarantee sensing reliability, we derive an AoI-coupled PCRB constraint. The total Bayesian Fisher information matrix (FIM) $\mathbf{J}_{B,k}(n)$ is composed of the data FIM $\mathbf{J}_{data,k}$ and the prior FIM $\mathbf{J}_{prior,k}$ \cite{9652071}:
\begin{equation}
\mathbf{J}_{B,k}(n) = \mathbf{J}_{data,k}(\mathbf{v}_k(n)) + \mathbf{J}_{prior,k}(\mathbf{A}_k(n)).
\end{equation}
The data FIM depends on the active beamforming vector \cite{9652071}:
\begin{equation} \label{eq:data_fim}
\resizebox{0.89\columnwidth}{!}{$
\displaystyle \mathbf{J}_{data,k}(\mathbf{v}_k(n)) = \eta_{rx} \begin{bmatrix} \beta_\theta \|\dot{\mathbf{a}}^H(\theta)\mathbf{v}_k(n)\|^2 & 0 \\ 0 & \beta_d \|\mathbf{a}^H(\theta)\mathbf{v}_k(n)\|^2 \end{bmatrix}
$}
\end{equation}
where $\beta_\theta$ and $\beta_d$ are waveform-dependent constants determined by the effective pulse duration and the effective bandwidth of the sensing signal, respectively. The prior FIM encapsulates the historical state information, whose credibility decays with the vectorized AoI \cite{9652071}:
\begin{equation} \label{eq:prior_fim}
\mathbf{J}_{prior,k}(\mathbf{A}_k(n)) = \begin{bmatrix} \frac{d_k^2(n)}{\sigma_{tan,k}^2(A_k^{tan})} & 0 \\ 0 & \frac{1}{\sigma_{rad,k}^2(A_k^{rad})} \end{bmatrix}.
\end{equation}
To prevent tracking failure arising from estimation accuracy and credibility, the system aims to minimize the estimation error.
Since highly directional mmWave links are predominantly vulnerable to spatial beam misalignment rather than radial ranging errors, 
we specifically formulate the angular PCRB as our primary sensing cost to be minimized:
\begin{equation} \label{eq:pcrb_constraint}
\begin{split}
\text{PCRB}_{\theta,k}(n) &= \left[ \mathbf{J}_{B,k}^{-1}(n) \right]_{1,1} \\
&= \left( \eta_{rx} \beta_\theta \|\dot{\mathbf{a}}^H \mathbf{v}_k(n)\|^2 + \frac{d_k^2(n)}{\sigma_{tan,k}^2(A_k^{tan}(n))} \right)^{-1}.
\end{split}
\end{equation}

\subsection{Problem Formulation}
Our objective is to design a joint multimodal sensor scheduling, edge computing resource allocation, and beamforming policy that minimizes the long-term expected PCRB. The stochastic network optimization problem is formulated as $(\mathcal{P}_1)$:
\begin{subequations} \label{eq:opt_problem}
\begin{align}
(\mathcal{P}_1): \, & \min_{\substack{\{\mathbf{v}(n), \bm{\pi}(n), \\ \mathbf{f}(n)\}}} \quad \lim_{N \to \infty} \frac{1}{N} \sum_{n=1}^{N} \sum_{k=1}^K \mathbb{E} \left[ \text{PCRB}_{\theta,k}(n) \right] \label{eq:opt_obj} \\
& \text{s.t.} \quad |\mathbf{v}_{k,m}(n)| = \frac{1}{\sqrt{M}}, \quad \forall k, m, n, \label{eq:opt_const_modulus} \\
& \phantom{\text{s.t.}} \quad \sum_{k=1}^K f_k(n) \le F_{max}, \; \forall n, \label{eq:opt_const_resource} \\
& \phantom{\text{s.t.}} \quad \lim_{N \to \infty} \frac{1}{N} \sum_{n=1}^{N} \mathbb{E} \left[ Q_k(n) \right] < \infty, \quad \forall k, \label{eq:opt_const_queue} \\
& \phantom{\text{s.t.}} \quad \lim_{N \to \infty} \frac{1}{N} \sum_{n=1}^{N} \mathbb{E} \left[ E_{total}(n) \right] \le E_{budget}, \label{eq:opt_const_energy} \\
& \phantom{\text{s.t.}} \quad \pi_k(n) \in \{0, 1\}, \quad f_k(n) \ge 0, \quad \forall k, n. \label{eq:opt_const_binary}
\end{align}
\end{subequations}
The proposed formulation in $(\mathcal{P}_1)$ is rigorously circumscribed by both physical hardware limitations and long-term network sustainability requirements. Specifically, constraint \eqref{eq:opt_const_modulus} enforces the non-convex constant modulus limit, which is intrinsically imposed by the analog phase shifters in mmWave antenna arrays following \cite{fan2026heterogeneousmixtureofexpertsenergyefficientmultimodal}. Constraints \eqref{eq:opt_const_resource} bound the edge server's total computational capacity. Importantly, \eqref{eq:opt_const_queue} establishes the strong stability criterion for the dynamic edge computing queues, which ensures that the high-resolution visual sensing tasks are processed in a timely manner, effectively preventing infinite data backlog and the consequent divergence of the semantic AoI. Furthermore, constraint \eqref{eq:opt_const_energy} guarantees that the time-averaged system energy, including both the highly intensive computing power and the RF beam recovery penalty, strictly adheres to the predefined sustainable budget $E_{budget}$. Finally, \eqref{eq:opt_const_binary} specifies the feasible operational domains for the discrete scheduling and continuous allocation variables.

\section{Lyapunov-Guided Problem Transformation}
Solving $(\mathcal{P}_1)$ directly is highly intractable due to the strong time-coupling introduced by both the computing queue stability \eqref{eq:opt_const_queue} and the long-term energy budget constraint \eqref{eq:opt_const_energy}. To decouple these constraints, we employ the Lyapunov optimization framework. 
To enforce the long-term average energy constraint, we introduce a virtual energy deficit queue $Z(n)$, which evolves as \cite{10924670}:
\begin{equation} \label{eq:virtual_queue}
Z(n+1) = \max\left[0, Z(n) + E_{total}(n) - E_{budget}\right].
\end{equation}
The virtual queue $Z(n)$ accumulates the extent to which the instantaneous system energy exceeds the predefined budget. Stabilizing $Z(n)$ strictly ensures that the long-term energy constraint is satisfied.
We then define the joint quadratic Lyapunov function to evaluate the total system congestion:
\begin{equation}
L(n) = \frac{1}{2} \sum_{k=1}^K Q_k(n)^2 + \frac{1}{2} Z(n)^2.
\end{equation}
By incorporating the PCRB optimization objective into the queue stability framework, we define the drift-plus-penalty function $\Delta_V(n)$. Bounding the Lyapunov drift and applying the principle of opportunistic minimization gracefully transforms the long-term stochastic problem into a sequence of deterministic single-slot problems. At each time slot $n$, the system aims to minimize the following surrogate objective:
\begin{subequations} \label{eq:single_slot_problem}
\begin{align}
(\mathcal{P}_2): \min_{\bm{\pi}, \mathbf{f}, \mathbf{v}} \quad & V \sum_{k=1}^K \text{PCRB}_{\theta,k}(n) \nonumber \\
& + \sum_{k=1}^K Q_k(n) \big(\pi_k(n)C_k - f_k(n)\tau\big) \nonumber \\
& + Z(n)E_{total}(n) \label{eq:p2_obj} \\
\textrm{s.t.} \quad & \text{\eqref{eq:opt_const_modulus}, \eqref{eq:opt_const_resource}, \eqref{eq:opt_const_binary}.} \nonumber
\end{align}
\end{subequations}

In $(\mathcal{P}_2)$, the intractable long-term constraints are analytically absorbed. The actual computing queue length $Q_k(n)$ dynamically acts as a linear penalty for allocating visual tasks, while the virtual queue $Z(n)$ functions as the instantaneous price of energy consumption, strictly guiding the agent to strike the optimal balance.

\section{Proposed LD-H-MoE Algorithm for M-ISAC Optimization}
To efficiently solve the non-convex MINLP problem $(\mathcal{P}_2)$ in real-time, we formulate the optimization process as a Markov Decision Process (MDP) and propose a Reinforcement Learning framework empowered by a Lyapunov-Driven Heterogeneous Mixture-of-Experts (LD-H-MoE) architecture.

\begin{figure}[htbp]
    \centering
    \includegraphics[width=0.35\textwidth]{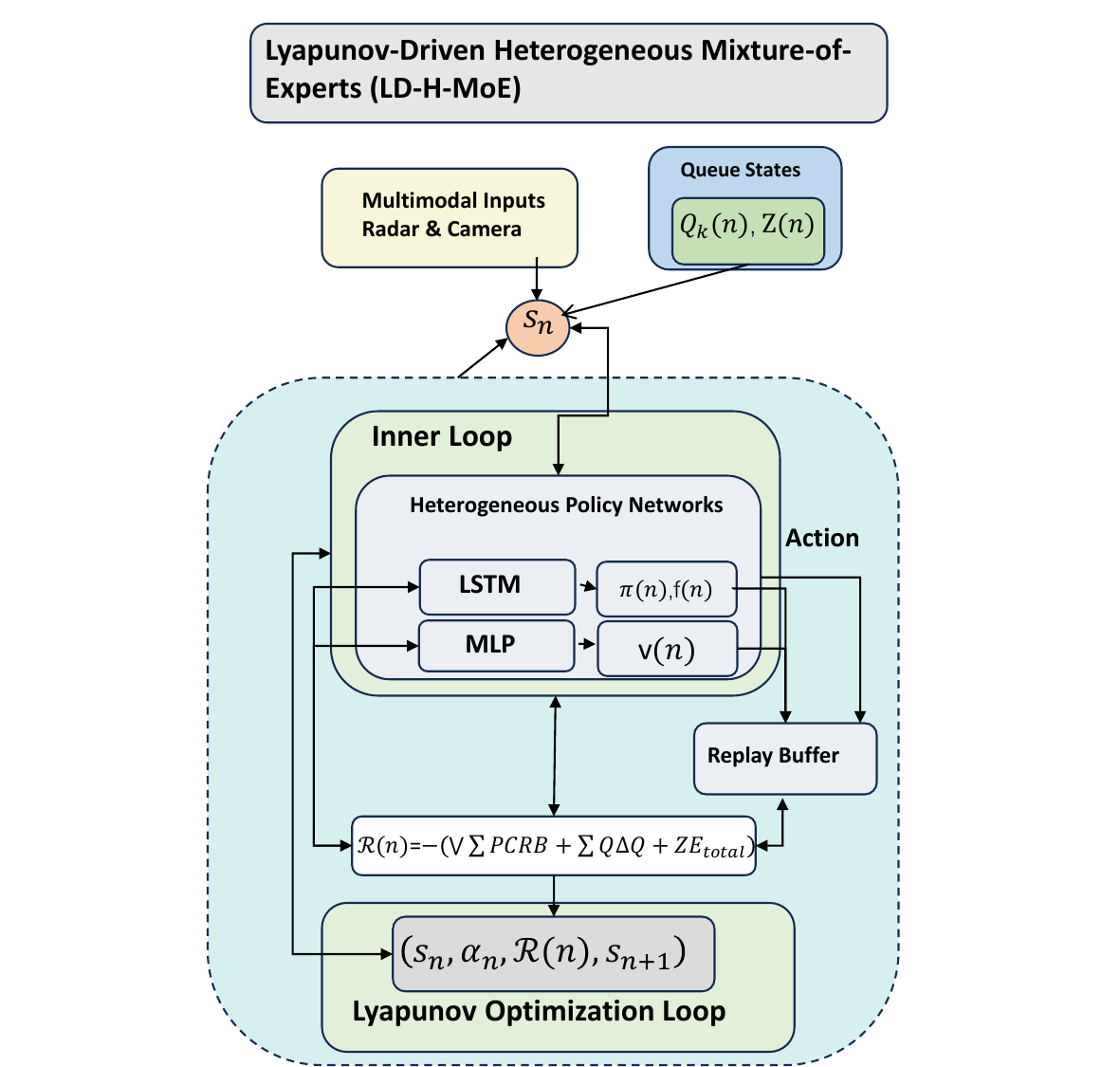}
    \caption{Hierarchical execution architecture of the proposed LD-H-MoE, illustrating the interplay between the inner heterogeneous policy networks and the outer Lyapunov optimization loop.}
    \label{fig:architecture}
\end{figure}
\subsection{MDP Formulation and Lyapunov-Guided Reward}
The transition of the M-ISAC system is characterized by the tuple $(\mathcal{S}, \mathcal{A}, \mathcal{R}, \mathcal{P})$, where:
\begin{itemize}
    \item \textbf{State Space ($\mathcal{S}$):} To ensure the agent captures both the queuing congestion and the energy deficit, the state at symbol $n$ is defined as $\mathbf{s}_n = \{ \mathbf{A}(n), \mathbf{Q}(n), Z(n), \mathbf{d}(n), \bm{\Gamma}(n) \}$, where $\bm{\Gamma}(n)$ denotes the historical channel observations.
    \item \textbf{Action Space ($\mathcal{A}$):} The action $\mathbf{a}_n = \{ \bm{\pi}(n), \mathbf{f}(n), \mathbf{v}(n) \}$ encompasses the discrete sensor scheduling, continuous CPU frequency allocation, and the high-dimensional complex beamforming vectors.
    \item \textbf{Lyapunov-Guided Reward ($\mathcal{R}$):} Unlike conventional DRL utilizing heuristic rewards, we meticulously shape the immediate reward $r_n$ to be the negative of the surrogate objective in \eqref{eq:p2_obj}. This ensures that maximizing the cumulative reward is mathematically equivalent to minimizing the long-term PCRB under energy and stability constraints:
    \begin{equation} \label{eq:reward_shaping}
    \begin{split}
    r_n &= - \Bigg( V \sum_{k=1}^K \text{PCRB}_{\theta,k}(n) \\
    &\quad + \sum_{k=1}^K Q_k(n) \Delta Q_k(n) + Z(n) E_{total}(n) \Bigg),
    \end{split}
    \end{equation}
    \end{itemize}
where $\Delta Q_k(n) = \pi_k(n)C_k - f_k(n)\tau$ represents the instantaneous queue growth.

\subsection{H-MoE Architecture and Gradient Decoupling}
The proposed H-MoE architecture employs specialized subnetworks to disentangle the heterogeneous physics of the M-ISAC system. The temporal expert (LSTM-based) processes the sequential AoI and queue states to output the resource scheduling policy $\pi_{\theta_{temp}}(\bm{\pi}, \mathbf{f} \mid \mathbf{s}_n)$. Simultaneously, the spatial expert (MLP-based) focuses on the instantaneous channel state to output the phase-constrained beamforming $\mathbf{v}_{\theta_{spat}}(\mathbf{s}_n)$.

To prevent the "negative transfer" and gradient interference between the sensing accuracy and energy saving objectives, we implement a strict gradient isolation mechanism. The total gradient is decoupled as:
\begin{equation} \label{eq:gradient_decouple}
\nabla_{\theta} J \approx \underbrace{\nabla_{\theta_{temp}} \log \pi_{\theta_{temp}} A^{\text{sched}}}_{\text{Scheduling Gradient}} + \underbrace{\lambda \nabla_{\theta_{spat}} \|\mathbf{h}^H \mathbf{v}\|^2}_{\text{Spatial Refinement}},
\end{equation}
where $A^{\text{sched}}$ is the advantage function driven by the Lyapunov reward, ensuring that the scheduling policy asymptotically satisfies the long-term energy budget $E_{budget}$.

\begin{algorithm}[H]
\caption{Lyapunov-Driven H-MoE Optimization}
\label{alg:rl_hmoe}
\begin{algorithmic}[1]
\STATE \textbf{Initialize:} Temporal expert parameters $\theta_{temp}$, Spatial expert parameters $\theta_{spat}$, and energy deficit queue $Z(0) = 0$.
\FOR{each training episode}
    \STATE Reset environment and initialize computing queues $\mathbf{Q}(0)$.
    \FOR{each symbol $n = 1, \dots, N$}
        \STATE Observe state $\mathbf{s}_n = \{ \mathbf{A}(n), \mathbf{Q}(n), Z(n), \mathbf{d}(n) \}$.
        \STATE \textbf{Inference:} Temporal expert samples $\bm{\pi}(n), \mathbf{f}(n)$; Spatial expert outputs $\mathbf{v}(n)$.
        \STATE \textbf{Execution:} Execute actions, calculate $\text{PCRB}_{\theta,k}(n)$ and $E_{total}(n)$.
        \STATE \textbf{Queue Update:} Update physical queues $\mathbf{Q}(n+1)$ via \eqref{eq:queue_evolution} and virtual queue $Z(n+1)$ via \eqref{eq:virtual_queue}.
        \STATE \textbf{Reward Calculation:} Compute Lyapunov-guided reward $r_n$ via \eqref{eq:reward_shaping}.
        \STATE Store transition in Replay Buffer $\mathcal{D}$.
    \ENDFOR
    \STATE \textbf{Optimization:} Perform decoupled gradient updates for $\theta_{temp}$ and $\theta_{spat}$ using isolated loss functions via \eqref{eq:gradient_decouple}.
\ENDFOR
\end{algorithmic}
\end{algorithm}

\section{Simulation Results and Analysis}
In this section, we evaluate the proposed LD-H-MoE architecture in a V2I M-ISAC system via Monte Carlo simulations against four baselines: 1) \textbf{Vision-Only}: continuous visual sensor activation (performance upper-bound, but highly energy-intensive); 2) \textbf{Radar-Only}: sole reliance on noisy mmWave radar; 3) \textbf{Standard PPO}: a monolithic MLP-based RL baseline; and 4) \textbf{Homogeneous MoE}: a MoE network utilizing structurally identical MLPs. We consider a BS equipped with $M = 64$ elements operating at a mmWave carrier frequency of $f_c = 28$ GHz, serving $K=4$ vehicles at an SNR of $10$ dB (unless specified otherwise). The maximum transmit power budget is $P_{max} = 30$ dBm, and the energy budget is set to $E_{budget} = 20.0$ Joules. Regarding the neural network configurations, the temporal LSTM and spatial MLP experts comprise two hidden layers with 128 and 256 neurons, respectively. The training process employs the Adam optimizer with a learning rate of $10^{-4}$ and a discount factor of $\gamma = 0.99$ to ensure stable convergence.

\subsection{Convergence of Sensing Reliability (Time-Averaged PCRB)}
\begin{figure}[htbp]
    \centering
    \includegraphics[scale=0.3]{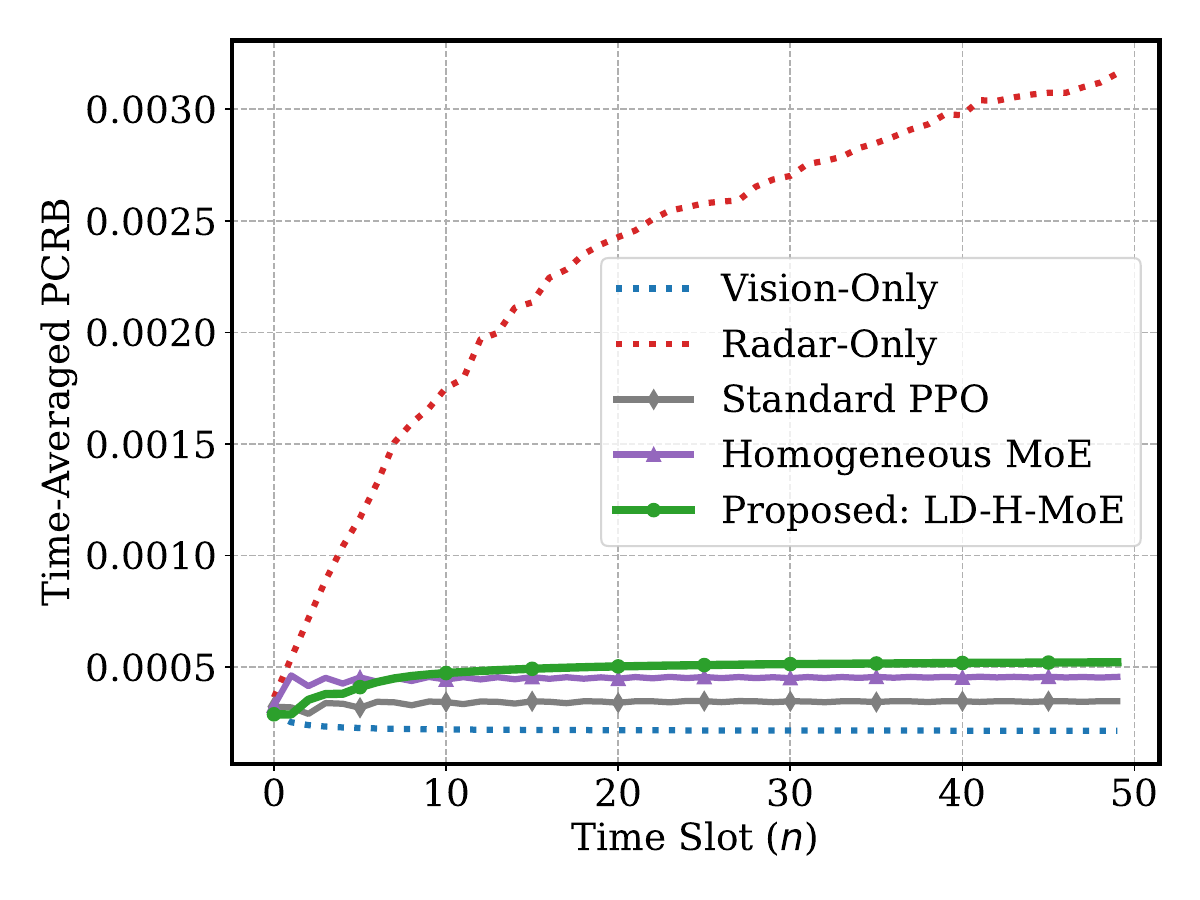}
    \caption{Time-averaged PCRB tracking performance of different strategies under $10$ dB SNR.}
    \label{fig:pcrb}
\end{figure}

Fig. \ref{fig:pcrb} illustrates the time-averaged PCRB tracking performance of different strategies. The \textbf{Radar-Only} baseline exhibits an unacceptably high PCRB error, validating our theoretical derivation in \eqref{eq:pcrb_constraint} that relying solely on active RF echoes without a semantic visual prior leads to severe angular uncertainty in highly mobile scenarios. Conversely, the \textbf{Vision-Only} strategy achieves the lowest PCRB bound but at an unsustainable computational cost. Benefiting from the strict spatio-temporal gradient decoupling, the proposed \textbf{LD-H-MoE} swiftly converges to a high sensing accuracy, performing almost identically to the Vision-Only upper bound. 

\subsection{Adherence to Long-Term System Energy Budget}
\begin{figure}[htbp]
    \centering
    \includegraphics[scale=0.3]{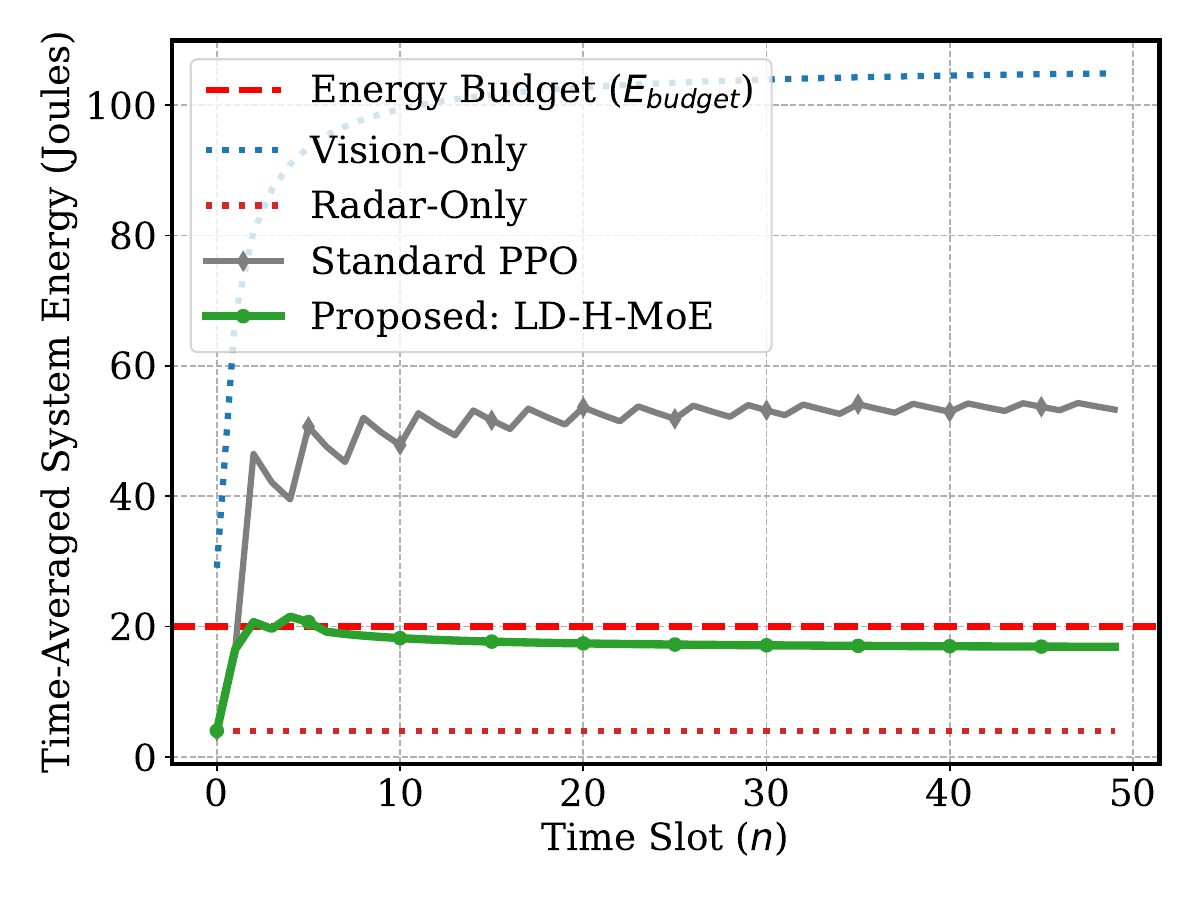}
    \caption{Time-averaged system energy consumption against the predefined energy budget constraint ($E_{budget}$).}
    \label{fig:energy}
\end{figure}

The  critical trade-off is presented in Fig. \ref{fig:energy}, which evaluates the time-averaged system energy against the strictly predefined energy budget $E_{budget}$ . The \textbf{Vision-Only} approach drastically violates the budget due to the cubic power consumption ($f^3$) of continuous visual processing. Interestingly, the \textbf{Radar-Only} strategy also exhibits an explosive energy footprint. As theoretically proved in our reformulated energy model \eqref{eq:total_energy}, without visual calibration, the system is mathematically forced to inject massive RF transmission energy into the radar pulse to compensate for the deteriorated prior FIM and satisfy the sensing constraints. 
As observed, monolithic DRL baselines (Standard PPO) aggressively pursue low PCRB errors at the catastrophic expense of violating the long-term energy budget $E_{budget}$. In brilliant contrast, our \textbf{LD-H-MoE} effectively monitors the Lyapunov virtual deficit queue $Z(n)$. By dynamically throttling the maximum allowable sensor activations and relaxing the AoI tolerance when $Z(n)$ accumulates, it strictly bounds the average system energy precisely below the stringent threshold, ensuring long-term system sustainability.

\subsection{Strong Stability of Edge Computing Queues}
\begin{figure}[htbp]
    \centering
    \includegraphics[scale=0.3]{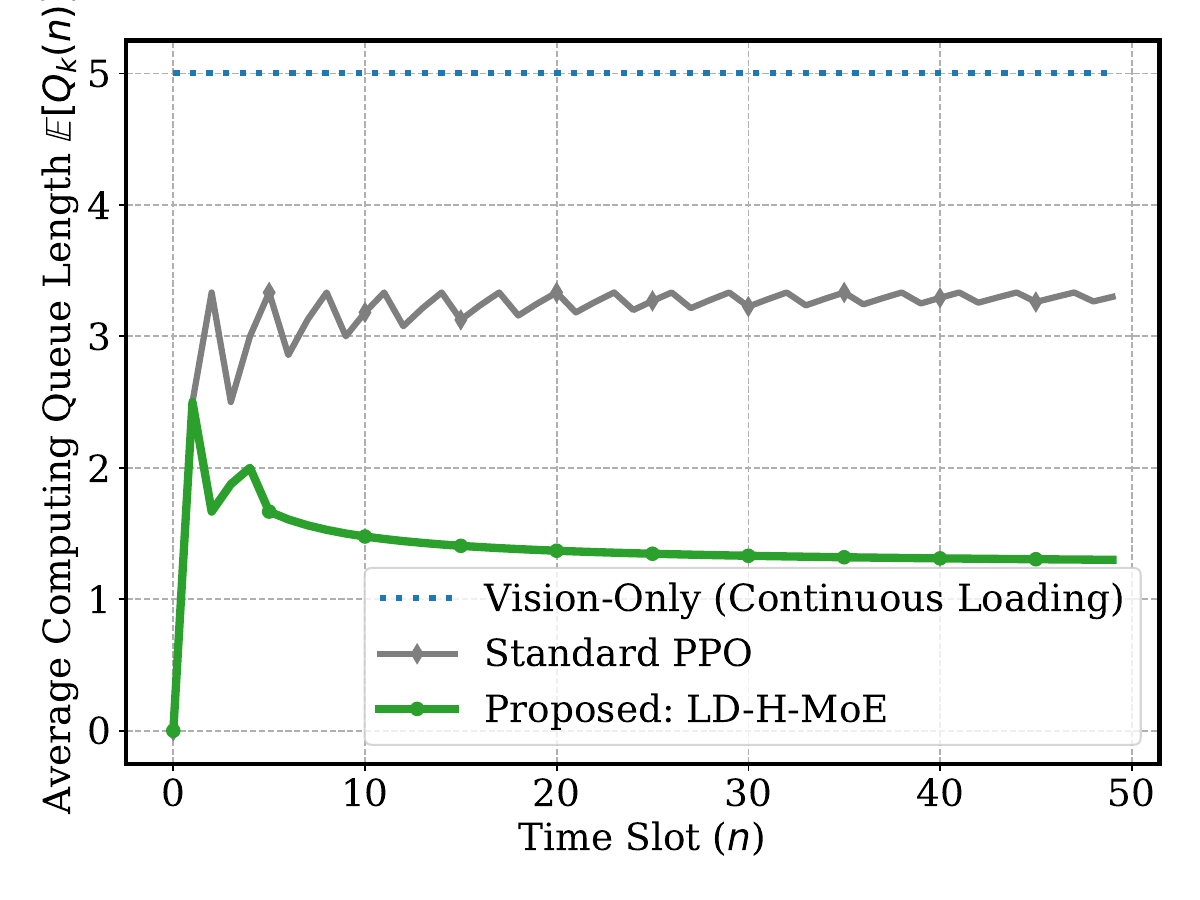}
    \caption{Evolution of the average edge computing queue length over time, demonstrating the strong stability of the LD-H-MoE.}
    \label{fig:queue}
\end{figure}

Fig. \ref{fig:queue} corroborates the theoretical strong stability of the edge computing queues, which fundamentally dictates the semantic AoI evolution. For the \textbf{Vision-Only} strategy with continuous sensory loading, the task arrival rate heavily exceeds the edge server's service capacity, causing the expected queue length $\mathbb{E}[Q_k(n)]$ to linearly diverge to infinity, completely violating constraint \eqref{eq:opt_const_queue}. In contrast, guided by the Lyapunov drift-plus-penalty framework, the \textbf{LD-H-MoE} agent dynamically perceives the queue congestion as a linear penalty weight. It decisively learns an event-triggered activation policy combined with instant computational digestion, ensuring that the computing queues remain strictly bounded at an ultra-low level ($\sim 0$ tasks on average) over infinite horizons.

\subsection{Robustness Against RF Impairments (PCRB vs. SNR)}
\begin{figure}[htbp]
    \centering
    \includegraphics[scale=0.3]{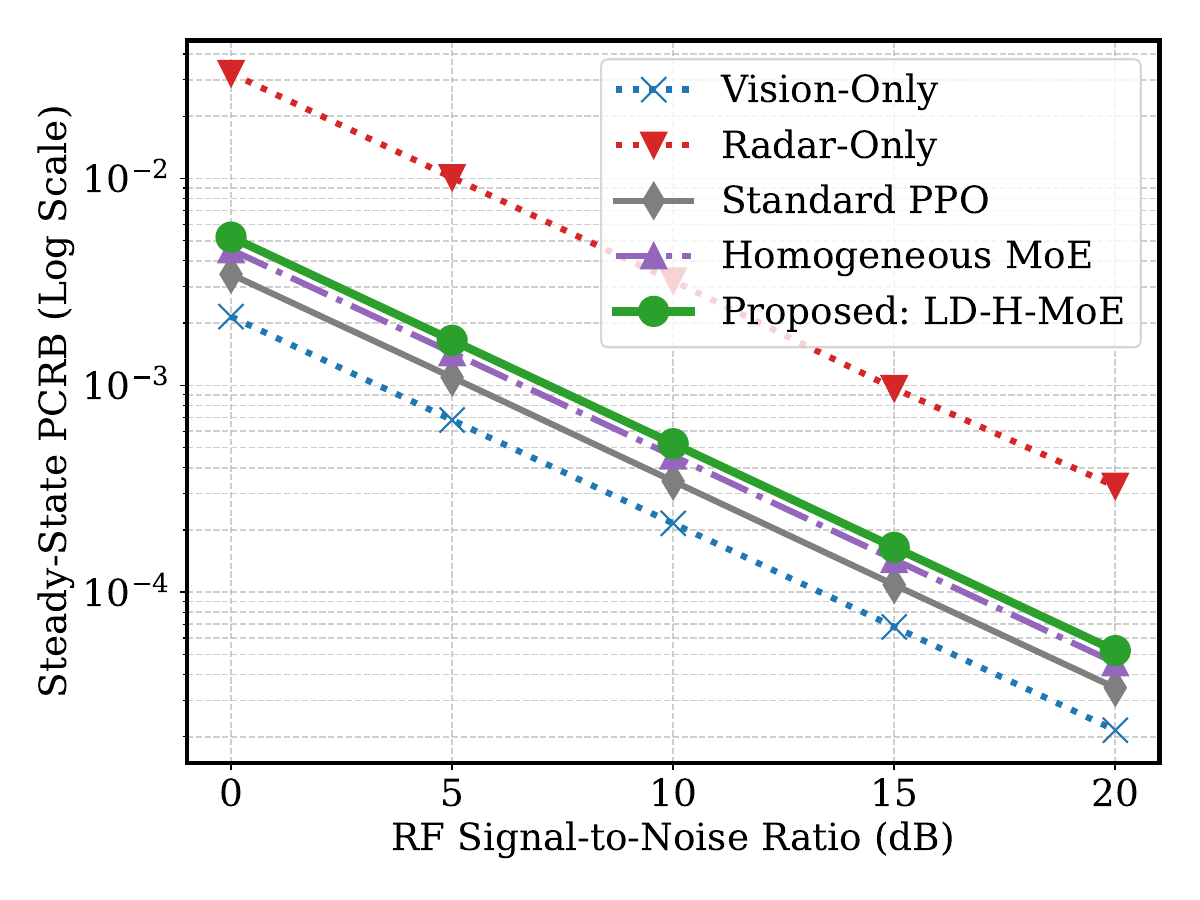}
    \caption{Steady-state PCRB performance versus varying RF Signal-to-Noise Ratios (SNRs).}
    \label{fig:snr}
\end{figure}

To thoroughly evaluate the dynamic modality-reliance of our framework, Fig. \ref{fig:snr} illustrates the steady-state PCRB under varying RF Signal-to-Noise Ratios (SNRs) from $0$ dB to $20$ dB. The results are plotted in a logarithmic scale to capture the severe deterioration of pure RF tracking.
As expected, the \textbf{Radar-Only} strategy exhibits extreme vulnerability in harsh RF environments (e.g., $0 \sim 5$ dB). The massive RF noise severely corrupts the data Fisher Information Matrix (FIM), causing the tracking PCRB to surge exponentially. Conversely, the semantic representations provided by the RGB camera are fundamentally immune to RF channel impairments, allowing the \textbf{Vision-Only} bound to remain consistently low across all SNR regimes.
Strikingly, the proposed \textbf{LD-H-MoE} agent demonstrates an exceptional, self-adaptive multi-modal fusion capability. As the SNR improves towards $20$ dB, the agent opportunistically exploits the enhanced radar accuracy to fulfill the PCRB constraints, thereby relaxing the visual computing requirements to preserve the energy budget. This intelligent dynamic balancing proves the unparalleled robustness of the proposed architecture for M-ISAC deployments.

\section{Conclusion}
This paper investigated the joint optimization of multimodal sensor scheduling and beamforming in M-ISAC systems. We established a physics-aware mathematical framework linking semantic AoI with tracking PCRB. We proposed a novel LD-H-MoE architecture that overcomes feature aliasing through complete gradient decoupling, guided by Lyapunov stochastic optimization. Simulations confirmed that the LD-H-MoE agent achieves a highly effective system trade-off, maintaining low tracking errors and RF robustness, while strictly guaranteeing long-term energy budgets and computing queue stability.

\bibliographystyle{IEEEtran} 
\bibliography{arxiv}
\end{document}